\documentclass{elsart}

\usepackage{amssymb}
\usepackage{units}
\usepackage{graphicx}

\begin{document}

\newcommand{\itl}{\tilde{\imath}}
\newcommand{\jtl}{\tilde{\jmath}}
\newcommand{\phlf}{\!+\!\half}
\newcommand{\mhlf}{\!-\!\half}
\newcommand{\pmhlf}{\!\pm\!\half}

\begin{frontmatter}

% Title , authors and addresses
  \title{Shearingbox-implementation for the central-upwind,
    constraint-transport MHD-code NIRVANA} \author{O. Gressel, U. Ziegler}
%\ead{uziegler@aip.de}
%\ead[url]{http://www.aip.de}
  \address{Astrophysikalisches Institut Potsdam, D-14482 Potsdam, Germany}


\begin{abstract}
% Text of abstract
We describe the implementation of the shearingbox approach into the
Godunov-type central-upwind/constraint-transport magnetohydrodynamics code
NIRVANA. This will allow for applications which require sheared-periodic
boundary conditions as typically used in local Cartesian simulations of
differentially rotating systems. We present the algorithm in detail and discuss
necessary modifications in the numerical fluxes in order to preserve conserved
quantities and to fulfill other analytical constraints as good as seem feasible
within the numerical scheme. We check the source terms which come with the
shearingbox formulation by investigating the conservation of the epicyclic mode
energy. We also perform more realistic simulations of the magneto-rotational
instability with initial zero-net-flux vertical magnetic field and compare the
obtained stresses and energetics with previous non-conservative results
exploring the same parameter regime.
\end{abstract}

\end{frontmatter}

% main text
\section{Introduction}
Simulations aiming to study the local behavior of (magneto-)hydrodynamic flows
usually invoke periodic boundary conditions.  This accounts for the
unboundedness of the flow and prevents the system from obtaining information
about its surface.  However, in the case of a shearing background flow
periodicity is not strictly applicable.  Typical scenarios for this include
differentially rotating gaseous accretion disks, which are encountered in
many astrophysical contexts. Such systems (if sufficiently ionized) are found to
be unstable to the so-called magneto-rotational instability (MRI) and have been
successfully modeled during the past decade (see \cite{BH98,B03} and references
therein). Considerable progress in the field has been made using the local
shearingbox approach, a detailed description of which can be found e.g.
in \cite{HGB95}.

The two key issues to be addressed in the design of shearingbox boundary
conditions are (1) the time-dependent shifted-periodic mapping of the
dependent variables of mass density, momentum, energy and, in case of
magnetohydrodynamic flows, the magnetic field and (2) the offset introduced in
the velocity component parallel to the shearing background.  The latter
represents the global shear across the domain which is forced constant in
time. Both of these features and the numerical treatment of the resulting source
terms
arising in the local approximation have profound implications on the accuracy
properties of the shearingbox advection/induction scheme.

In this paper we will deal with these implications in detail. First, section 2
gives a compact recollection of basic facts about the MHD scheme in NIRVANA
necessary for a understanding of the following shearingbox implementation
which is then described in Section 3. Section 4 presents test cases to
validate our approach.

\section{Review of the MHD scheme}

Although the NIRVANA code can account for dissipative effects (viscosity,
diffusion, thermal heat conduction), multi-scale phenomena (using adaptive
mesh refinement, see Ziegler \cite{Z05}) as well as self-gravity
(multigrid-type Poisson solver, see Ziegler \cite{Z05b}), we want to focus
here on ideal MHD in three space dimensions described by the following set of
coupled equations

\begin{eqnarray}
\frac{\partial {\bf u}}{\partial t}+\frac{\partial {\bf f}^x}{\partial x}
+\frac{\partial {\bf f}^y}{\partial y}+\frac{\partial {\bf f}^z}{\partial z}
& = & {\bf S}({\bf u})\,,\label{eq:hydro} \\
\frac{\partial {\bf B}}{\partial t} +\nabla\!\times\!{\bf E} 
& = & 0\,, \label{eq:induction}
\end{eqnarray}

where ${\bf u}=(\varrho, m_x,m_y,m_z,e)^{\top}$. The source term ${\bf
S}$ arises from non-inertial forces in the local frame
(rotating with $\bf \Omega$ relative to the inertial frame) and will be
discussed in detail in section \ref{sec:source}.
The flux functions ${\bf f}^x$, ${\bf f}^y$. ${\bf f}^z$ are defined as:

\begin{eqnarray}
{\bf f}^x & = &
\left(\begin{array}{c}
m_x \\[-1ex] 
m_xv_x+p+\half{\bf B}^2 - B_xB_x \\[-1ex]
m_yv_x-B_yB_x \\[-1ex]
m_zv_x-B_zB_x \\[-1ex]
(e+p+\half{\bf B}^2)\,v_x-({\bf v}\!\cdot\!{\bf B})B_x
\end{array}\right)\,,\label{eq:flux_x} \\
{\bf f}^y & = &
\left(\begin{array}{c}
m_y \\[-1ex]
m_xv_y-B_xB_x \\[-1ex]
m_yv_y+p+\half{\bf B}^2 - B_yB_y \\[-1ex]
m_zv_y - B_zB_y \\[-1ex]
(e+p+\half{\bf B}^2)\,v_y-({\bf v}\!\cdot\!{\bf B})B_y
\end{array}\right)\,, \\
{\bf f}^z & = &
\left(\begin{array}{c}
m_z \\[-1ex]
m_xv_z - B_xB_z \\[-1ex]
m_yv_z - B_yB_z \\[-1ex]
m_zv_z+p+\half{\bf B}^2 - B_zB_z \\[-1ex]
(e+p+\half{\bf B}^2)\,v_z-({\bf v}\!\cdot\!{\bf B})B_z
\end{array}\right)\,.
\end{eqnarray}

We adopt a notation where the magnetic permeability is set to unity. The
equations are to be supplemented by the zero-divergence constraint
$\nabla\!\cdot\!{\bf B}=0$ for the magnetic field {\bf B} and an equation of
state $e=\epsilon+\half\varrho{\bf v}^2+\half{\bf B}^2 $ with
$\epsilon=p/(\gamma -1)$. Other symbols have their usual meaning with
$\varrho$ denoting the mass density, $e$ the total energy density, $\epsilon$
the thermal energy density, $p$ the pressure, ${\bf v}$ the velocity, ${\bf
  m}=\varrho{\bf v}$ the momentum, ${\bf E}=-{\bf v}\times{\bf B}$ the
electric field and $\gamma$ the ratio of specific heats.

The numerical method used to solve the above system is a hybrid scheme
combining the second-order version of the Godunov-type central-upwind scheme
of Kurganov et al. \cite{KNP01} with the constraint transport (CT) technique for
divergence free magnetic field evolution (see e.g. Evans \& Hawley
\cite{EH88}, T\'oth \cite{T00}). The original description of this hybrid
approach can be found in Ziegler \cite{Z04}. Here, we compactly summarize the
relevant knowledge necessary for constructing the shearingbox-implementation.

Let the computational domain be subdivided in uniform Cartesian cells
$C_{i,j,k}\!=\![x_{i\mhlf},x_{i\phlf}] \times [y_{j\mhlf},y_{j\phlf}]\times
[z_{k\mhlf},z_{k\phlf}]$ with center $(x_i,y_j,z_k)$ and spacings $\delta
x$, $\delta y$ and $\delta z$.  Then, the scheme for Eq.~(\ref{eq:hydro})
in semi-discrete form reads

\begin{eqnarray}
\frac{d}{dt}\overline{\bf u}_{i,j,k}(t) & = & 
-\frac{{\bf F}^x_{i\phlf,j,k}(t)-{\bf F}^x_{i\mhlf,j,k}(t)}{\delta x}
-\frac{{\bf F}^y_{i,j\phlf,k}(t)-{\bf F}^y_{i,j\mhlf,k}(t)}{\delta y}
\nonumber \\&& 
-\frac{{\bf F}^z_{i,j,k\phlf}(t)-{\bf F}^z_{i,j,k\mhlf}(t)}{\delta z}
+{\bf S}(\overline{\bf u}_{i,j,k}(t)) \label{eq:hydro_sd}
\end{eqnarray}

where the cell-centered $\overline{\bf u}_{i,j,k}$ is an approximation to the 
volume-averaged state ${\bf u}$ in cell $C_{i,j,k}$ and the numerical fluxes are given by:

\begin{eqnarray}
{\bf F}^x_{i\phlf,j,k} & = & \frac{a^+{\bf f}^x({\bf w}^E_{i,j,k})
-a^-{\bf f}^x({\bf w}^W_{i+1,j,k})}{a^+-a^-}
+\frac{a^+a^-\left({\bf u}^W_{i+1,j,k}-{\bf u}^E_{i,j,k}\right)}{a^+-a^-}\,,
\label{eq:Flux_x}\\
{\bf F}^y_{i,j\phlf,k} & = & \frac{b^+{\bf f}^y({\bf w}^N_{i,j,k})
-b^-{\bf f}^y({\bf w}^S_{i,j+1,k})}{b^+-b^-}
+\frac{b^+b^-\left({\bf u}^S_{i,j+1,k}-{\bf u}^N_{i,j,k}\right)}{b^+-b^-}\,,\\
{\bf F}^z_{i,j,k\phlf} & = & \frac{c^+{\bf f}^z({\bf w}^T_{i,j,k})
-c^-{\bf f}^z({\bf w}^B_{i,j,k+1})}{c^+-c^-}
+\frac{c^+c^-\left({\bf u}^B_{i,j,k+1}-{\bf u}^T_{i,j,k}\right)}{c^+-c^-}
\end{eqnarray}

with ${\bf w}\!=\!({\bf u},{\bf B})^{\top}$. We apply the third-order 
strong-stability-preserving Runge-Kutta scheme in \cite{GST01}
to integrate Eq.~(\ref{eq:hydro_sd}) in time (see also Eq.~(\ref{eq:rk3})).
At any stage in the Runge-Kutta scheme the evaluation of the flux functions at cell faces
requires point values for the variables left/right to a cell interface
which are obtained by piecewise linear reconstruction from 
${\overline{\bf w}_{i,j,k}}$ applying the TVD slope-limiter of van Leer
\cite{vL77}.
These point values are characterized in the following by superscripts
W at cell interface $x=x_{i\mhlf}$, E at $x=x_{i\phlf}$, S at $y=y_{j\mhlf}$, N
at $y=y_{j\phlf}$, B at $z=z_{k\mhlf}$ and T at $z=z_{k\phlf}$, respectively.
With such notation
${\bf w}^W_{i+1,j,k}$ stands for the reconstructed value within cell
$C_{i+1,j,k}$ at $x=x_{i\phlf}$ and ${\bf w}^E_{i,j,k}$ for the value within
cell $C_{i,j,k}$ at the {\em same\/} position. 
In contrast to the {\it cell-centered} ${\overline{\bf u}_{i,j,k}}$  we use staggering
for the magnetic field i.e. the {\it face-centered} finite-volume approximations to the 
face-averaged components are 
$\overline{B}_{x\vert i\mhlf,j,k}$, $\overline{B}_{y\vert i,j\mhlf,k}$ and
$\overline{B}_{z\vert i,j,k\mhlf}$, respectively.
It follows that in the reconstruction of {\bf B} at E,W position the x-component is
already properly located, i.e. $(B_x)_{i,j,k}^{E,W}=\overline{B}_{x\vert i\pmhlf,j,k}$,
whereas other components are first reconstructed in $x$-direction and then 
spatially averaged in the transverse direction. Reconstruction of ${\bf B}$ at the
other locations N,S,B,T is done in analogous fashion.

The quantities $a^{\pm}$ in the flux formulae define
the maximum (plus sign) respective minimum (minus sign)
wave-propagation-direction-sensitive speed at the interface $x_{i\phlf}$,
i.e.

\begin{eqnarray}
a^+&=&\max\{(v_x+c_{\mathrm{f}})^W_{i+1,j,k},\,
            (v_x+c_{\mathrm{f}})^E_{i,j,k},\,0\}\,, \nonumber \\
a^-&=&\min\{(v_x-c_{\mathrm{f}})^W_{i+1,j,k},\,
            (v_x-c_{\mathrm{f}})^E_{i,j,k},\,0\}\,, \nonumber
\end{eqnarray} 

where $c_{\mathrm{f}}=(c_S^2+c_A^2)^{1/2}$ is an upper limit for the fast
magnetosonic speed including the sound speed
$c_{\mathrm{s}}=(\gamma p/\varrho)^{1/2}$ and Alfv\'en-speed $c_A=({\bf
  B}^2/\varrho)^{1/2}$.
The corresponding speeds in the $y$($z$)-direction are denoted by $b^{\pm}$
($c^{\pm}$).

The above method can be formally applied to the induction
equation~(\ref{eq:induction}). Writing the curl of the electric field as
divergence of an antisymmetric tensor

\[
-\nabla\times{\bf E}=\nabla\!\cdot\!\left(
\begin{array}{ccc}
0 & E_z & -E_y \\[-1ex]
-E_z & 0 & E_x \\[-1ex]
E_y & -E_x & 0
\end{array} \right)\,,
\]

and utilizing the abbreviations
${\mbox{\boldmath$\varepsilon$}}^x=(0,-E_z,E_y)^{\top}$,
${\mbox{\boldmath$\varepsilon$}}^y=(E_z,0,-E_x)^{\top}$, and
${\mbox{\boldmath$\varepsilon$}}^z=(-E_y,E_x,0)^{\top}$, in analogy to Eqs.
(7)--(9) one can derive electric field fluxes

\begin{eqnarray}
{\bf G}^x_{i\phlf,j,k} & = & 
\frac{a^+{\mbox{\boldmath$\varepsilon$}}^x({\bf w}^E_{i,j,k})
-a^-{\mbox{\boldmath$\varepsilon$}}^x({\bf w}^W_{i+1,j,k})}{a^+-a^-}
+\frac{a^+a^-\left({\bf B}^W_{i+1,j,k}-{\bf B}^E_{i,j,k}\right)}{a^+-a^-}
\,,\\
{\bf G}^y_{i,j\phlf,k} & = & 
\frac{b^+{\mbox{\boldmath$\varepsilon$}}^y({\bf w}^N_{i,j,k})
-b^-{\mbox{\boldmath$\varepsilon$}}^y({\bf w}^S_{i,j+1,k})}{b^+-b^-}
+\frac{b^+b^-\left({\bf B}^S_{i,j+1,k}-{\bf B}^N_{i,j,k}\right)}{b^+-b^-}
\,,\\
{\bf G}^z_{i,j,k\phlf} & = & 
\frac{c^+{\mbox{\boldmath$\varepsilon$}}^z({\bf w}^T_{i,j,k})
-c^-{\mbox{\boldmath$\varepsilon$}}^z({\bf w}^B_{i,j,k+1})}{c^+-c^-}
+\frac{c^+c^-\left({\bf B}^B_{i,j,k+1}-{\bf B}^T_{i,j,k}\right)}{c^+-c^-}\,.
\end{eqnarray}

These are, like the ${\bf F}$-fluxes, defined at cell faces.
As before, at any stage in the Runge-Kutta scheme, the electric field function
${\mbox{\boldmath$\varepsilon$}}^x$  (${\mbox{\boldmath$\varepsilon$}}^y$,
${\mbox{\boldmath$\varepsilon$}}^z$) have to be evaluated at E,W (N,S,T,B) cell
interfaces using reconstructed data, e.g.,
${\varepsilon}^x_y({\bf w}^E_{i,j,k})=(-E_z)^E_{i,j,k}
=-(v_x)^E_{i,j,k}(B_y)^E_{i,j,k}+(v_y)^E_{i,j,k}(B_x)^E_{i,j,k}$.
Our CT scheme for the staggered magnetic field components finally reads in
semi-discrete form

\begin{eqnarray}
\frac{d}{dt}\overline{B}_{x\vert i\mhlf,j,k} & = &
-\frac{\overline{E}_{z\vert i\mhlf,j\phlf,k}
-\overline{E}_{z\vert i\mhlf,j\mhlf,k}}{\delta y}
+\frac{\overline{E}_{y\vert i\mhlf,j,k\phlf}
-\overline{E}_{y\vert i\mhlf,j,k\mhlf}}{\delta z}\,, \label{eq:ind_x_sd} \\
\frac{d}{dt}\overline{B}_{y\vert i,j\mhlf,k} & = &
+\frac{\overline{E}_{z\vert i\phlf,j\mhlf,k}
-\overline{E}_{z\vert i\mhlf,j\mhlf,k}}{\delta x}
-\frac{\overline{E}_{x\vert i,j\mhlf,k\phlf}
-\overline{E}_{x\vert i,j\mhlf,k\mhlf}}{\delta z}\,, \label{eq:ind_y_sd} \\
\frac{d}{dt}\overline{B}_{z\vert i,j,k\mhlf} & = &
-\frac{\overline{E}_{y\vert i\phlf,j,k\mhlf}
-\overline{E}_{y\vert i\mhlf,j,k\mhlf}}{\delta x}
+\frac{\overline{E}_{x\vert i,j\phlf,k\mhlf}
-\overline{E}_{x\vert i,j\mhlf,k\mhlf}}{\delta y}\,  \label{eq:ind_z_sd}
\end{eqnarray}

where $\overline{E}_{x\vert i,j\mhlf,k\mhlf}$, $\overline{E}_{y\vert
  i\mhlf,j,k\mhlf}$, $\overline{E}_{z\vert i\mhlf,j\mhlf,k}$ are
{\it edge-centered} approximations to the edge-averaged electric field components.
These are computed by composition of the electric field
fluxes obtained from the Godunov-central scheme, i.e.,

\begin{eqnarray}
\overline{E}_{x\vert i,j\mhlf,k\mhlf} & = &
\frac{1}{4}( -{\bf G}^y_{z\vert i,j\mhlf,k}-{\bf G}^y_{z\vert i,j\mhlf,k-1}
+{\bf G}^z_{y\vert i,j,k\mhlf}+{\bf G}^z_{y\vert i,j-1,k\mhlf})\,,  \\
\overline{E}_{y\vert i\mhlf,j,k\mhlf} & = &
\frac{1}{4}( +{\bf G}^x_{z\vert i\mhlf,j,k}+{\bf G}^x_{z\vert i\mhlf,j,k-1}
-{\bf G}^z_{x\vert i,j,k\mhlf}-{\bf G}^z_{x\vert i-1,j,k\mhlf})\,, \\
\overline{E}_{z\vert i\mhlf,j\mhlf,k} & = &
\frac{1}{4}( -{\bf G}^x_{y\vert i\mhlf,j,k}-{\bf G}^x_{y\vert i\mhlf,j-1,k}
+{\bf G}^y_{x\vert i,j\mhlf,k}+{\bf G}^y_{x\vert i-1,j\mhlf,k})\,.
\end{eqnarray}
It is now easy to show from (\ref{eq:ind_x_sd})--(\ref{eq:ind_z_sd}) that
\begin{eqnarray}
\frac{d}{dt}\left(\nabla\!\cdot\!\overline{\bf B}\right)_{i,j,k} & =
\frac{d}{dt} & \left(
\frac{\overline{B}_{x\vert i\phlf,j,k}\!-\!\overline{B}_{x\vert
i\mhlf,j,k}}{\delta x} \right. \nonumber \\ \,&\, & \left.
+\frac{\overline{B}_{y\vert i,j\phlf,k}\!-\!\overline{B}_{y\vert
i,j\mhlf,k}}{\delta y}
+\frac{\overline{B}_{z\vert i,j,k\phlf}\!-\!\overline{B}_{z\vert
i,j,k\mhlf}}{\delta z}
\right)=0\,. \nonumber
\end{eqnarray}

Thus, if 
$(\nabla\!\cdot\!\overline{\bf B})_{i,j,k}=0$ initially, the system evolves divergence-free 
since this condition is fulfilled after each stage in the multi-stage Runge-Kutta
scheme.

The full system of ODEs~(\ref{eq:hydro_sd}),~%
(\ref{eq:ind_x_sd})--(\ref{eq:ind_z_sd}) including the source term is solved 
with the strong-stability-preserving third-order Runge-Kutta integration scheme given by
(dropping cell indices and writing the rhs of Eq. \ref{eq:hydro_sd} as 
${\bf L}_F + {\bf S}$
and (\ref{eq:ind_x_sd})--(\ref{eq:ind_z_sd}) compactely as 
$\frac{d}{dt}{\overline{\bf B}}={\bf L}_E$)
\begin{eqnarray}
{\overline{\bf u}}^{(1)} & = & {\overline{\bf u}}^n+\delta t({\bf L}^n_F+{\bf S}^n)\nonumber\\
{\overline{\bf B}}^{(1)} & = & {\overline{\bf B}}^n+\delta t{\bf L}^n_E\,, \nonumber\\
{\overline{\bf u}}^{(2)} & = & \frac{3}{4}{\overline{\bf u}}^n
+\frac{1}{4}{\overline{\bf u}}^{(1)}
+\frac{1}{4}\delta t({\bf L}^{(1)}_F+{\bf S}^{(1)}) \nonumber\\
{\overline{\bf B}}^{(2)} & = & \frac{3}{4}{\overline{\bf B}}^n
+\frac{1}{4}{\overline{\bf B}}^{(1)}
+\frac{1}{4}\delta t{\bf L}^{(1)}_E \,,\nonumber\\
{\overline{\bf u}}^{n+1} & = & \frac{1}{3}{\overline{\bf u}}^n
+\frac{2}{3}{\overline{\bf u}}^{(2)}
+\frac{2}{3}\delta t({\bf L}^{(2)}_F+{\bf S}^{(2)}) \nonumber\\
{\overline{\bf B}}^{n+1} & = & \frac{1}{3}{\overline{\bf B}}^n
+\frac{2}{3}{\overline{\bf B}}^{(2)}
+\frac{2}{3}\delta t{\bf L}^{(2)}_E
\label{eq:rk3}
\end{eqnarray}
with the time-step $\delta t=t^{n+1}-t^n$ restricted by the CFL condition.

\section{Shearingbox implementation}

\begin{figure}
  \centering \includegraphics[width=11.5cm]{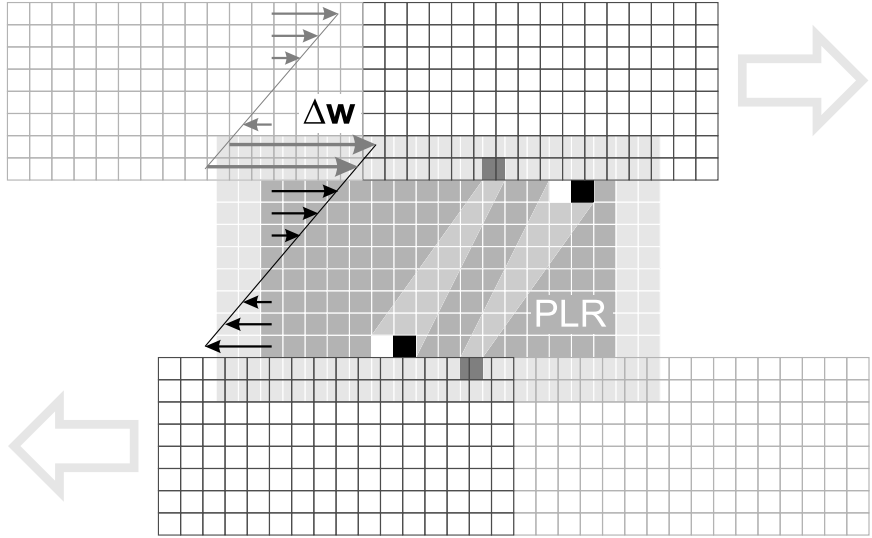}
\caption{Mapping of ghost zone values for a 2D shearingsheet. Reconstruction is
  indicated by highlighted cells. Arrows illustrate the need for a global
  velocity offset to match the velocity profile of the neighboring domain.}
\label{fig:shearing}
\end{figure}

Let us now consider a magnetohydrodynamic flow with a background profile of
the form ${\bf v}=(0,v_y=s x,0)$, with (linear) shear-parameter $s$.  The
computational domain is a Cartesian box with dimensions $L_x$, $L_y$, $L_z$
and periodic boundaries in $y-$ and $z-$direction. To account for the
background shear we introduce shearingbox boundary conditions in
$x-$direction. These can be expressed mathematically in the
form\footnote{Position arguments are being suppressed for extra variables in
  eqs.~(\ref{eq:map_my}), (\ref{eq:map_e}).} (e.g. \cite{HGB95}):

\begin{eqnarray}
& &
f(x,y,z) \mapsto f(x\pm L_x, y\mp w t, z), \qquad\qquad
f \in \left\lbrace \varrho, m_x, m_z, \epsilon \right\rbrace \label{eq:map_f}\\
& & m_y(x,y,z) \mapsto m_y(x\pm L_x, y\mp w t, z)\,\mp \varrho w,
\label{eq:map_my}
\end{eqnarray}

where $w=-s L_x$ represents the global velocity-offset across the box, as can
be seen in Figure~\ref{fig:shearing}. Because NIRVANA being a conservative
code evolves the total energy (rather than the internal energy $\epsilon$) we
supplement the corresponding relation for the total energy density

\begin{eqnarray}
& & e(x,y,z) \mapsto e(x\pm L_x, y\mp w t, z)\,\mp m_y w\, +\half
\varrho w^2 \label{eq:map_e},
\end{eqnarray}

which can be easily derived by separating the $m_y^2/(2\varrho)$ part from the
total energy and then using~(\ref{eq:map_my}).

As the $y$-coordinate of above mappings varies continously in time, there is
some kind of interpolation necessary to map ghostzone values on a finite grid
(see e.g. \cite{HGB95} for details). For our implementation, the same
piecewise linear reconstruction is used as for the numerical scheme.

Due to the shifted periodicity and the additional interpolation, ghost cells
on opposing sides of the domain along with their adjoint regular zones are not
redundant in the same sense as for strictly periodic boundary conditions. Let
us elucidate on that: In the case of regular periodicity every pair of cell
and ghost cell (separated by the domain boundary) has an identical counterpart
at the opposite boundary. Therefore the total flux across the interface is
conserved to machine accuracy.

For shifted periodicity at a given time $t$ this does no longer hold. Thus the
additional truncation error connected to the interpolation can lead to the
buildup of considerable deviations in conserved quantities, as is shown in
section~\ref{sec:adv_test}. In the following we want to suggest a method to
correct for this. Since strict conservation only applies as long as there is no
source on the rhs of (\ref{eq:hydro}), in the local shearingbox approach
with source term (\ref{eq:source_m})/(\ref{eq:source_e}) only the total
z-momentum and mass are exactly conserved. We note that it is nevertheless
desirable to preserve all variables as good as possible under advection.

\subsection{Conservation of hydrodynamic variables}
As the numerical fluxes are nonlinear functions of the conserved quantities
any form of interpolation for the ghost cells will lead to some small
inconsistency in the fluxes. This can be avoided by matching the computed
$x$-fluxes at the sheared domain boundaries. It is straightforward to map
fluxes not containing $m_y$. The quantities related to those fluxes, i.e. mass
density, $x$- and $z$-momentum are then conserved to machine precision with
respect to advection.  For the $y$-momentum flux and total energy flux, applying
the mappings~(\ref{eq:map_my})/(\ref{eq:map_e}) to the third and fifth
component of the flux function~(\ref{eq:flux_x}) yields:

\begin{eqnarray}
f^x(m_y) & \,\mapsto\, & f^x(m_y) \mp f^x(\varrho)\, w\;,
\label{eq:map_fmy} \\
f^x(e)   & \,\mapsto\, & f^x(e)   \mp f^x(m_y)\, w +\half f^x(\varrho)\, w^2\;,
\label{eq:map_fe}
\end{eqnarray}

in nice analogy to the original mappings. We want to point out that these
relations consistently include the magnetic part of the fluxes due to the
Lorentz-force, although this is not directly visible in the above notation.
Also, the modifications to the fluxes can be solely expressed in terms of
fluxes and the velocity offset $w$.

The recipe can then directly be carried over to the numerical
fluxes~(\ref{eq:Flux_x}): while the first term is just a linear combination of
the flux-function (evaluated at the adequate positions) one has to apply
(\ref{eq:map_my})/(\ref{eq:map_e}) once again for the second term, which then
gives:

\begin{eqnarray}
F^x_{i\phlf,j,k}(m_y) &\, = \,& \widehat{F}^x_{\itl\phlf,\jtl,k}(m_y)
\;\mp\;\widehat{F}^x_{\itl\phlf,\jtl,k}(\varrho)\, w\;,\label{eq:map_Fmy} \\
F^x_{i\phlf,j,k}(e) &\, = \,&\widehat{F}^x_{\itl\phlf,\jtl,k}(e) 
\;\mp\;\widehat{F}^x_{\itl\phlf,\jtl,k}(m_y)\, w
\;+\;\half \widehat{F}^x_{\itl\phlf,\jtl,k}(\varrho)\,w^2\;,\label{eq:map_Fe}
\end{eqnarray}

where the hat stands for the piecewise linear interpolation procedure used, and
tilde marks the corresponding indices of the zones to map from.

Relation~(\ref{eq:map_fe}) expresses the fact that the total energy within the
shearingbox is not conserved but can be altered by angular momentum transport
through the radial boundaries. In integral formulation this corresponds to
Eq.~(8) (cast into our notation) of Hawley et al. \cite{HGB95}:

\begin{equation}
\frac{\partial \Gamma}{\partial t}= w \int_{\partial X}{\mathrm{dy\,dz} \left(
\varrho
v_x \delta v_y - B_x B_y \right) }\label{eq:ene_cons}
\end{equation}

where $\Gamma$ is the volume integral of total energy and $\delta v_y=v_y+sx$
expresses the perturbed velocity.\footnote{Expanding $\delta v_y$ yields the
  additional $w^2$ term in~(\ref{eq:map_fe}).} When correcting the energy- and
momentum-fluxes according to Eqs.~(\ref{eq:map_fmy}), (\ref{eq:map_fe})
one can satisfy this property more accurately which allows to trace the
detailed evolution of energetics, e.g. in MRI-simulations.

\subsection{Conservation of magnetic flux}\label{sec:cons_magn}
Applying Gau\ss{}' theorem to the integral form of the induction equation, one
can show that the azimuthal field (due to the shear) grows linearly with the
net radial magnetic field-flux through the radial boundaries:

\begin{equation}
\frac{\partial \left<{\bf B}\right>}{\partial t} =
-\frac{w}{V}\; \hat{\bf y} \int_{\partial X}{\mathrm{dy\,dz} B_x}.
\end{equation}

For zero net radial field the mean magnetic flux through the shearingbox is
conserved. Our implementation satisfies this condition to machine accuracy for
the $x$- and $z$-component of the magnetic field and to truncation error for
the $y$-component (see section~\ref{sec:adv_test}). This corresponds to the
amount of precision reported by
\cite{HGB95,GS05}.

The seminal paper of Hawley et. al \cite{HGB95} discusses this topic rather
briefly, mentioning that mapping the electromotive forces at the sheared
interfaces conserves the vertical field to roundoff error. Furthermore, the
authors state that the spurious vertical field, that arises otherwise, is
negligible as the associated MRI growth rates are unresolved. Considering the
differences between MRI-simulations with and without zero net vertical field
(see e.g. \cite{GS05}) leaves this somewhat questionable in the sense that a 
spurious field might lead to an overestimation of magnetic stresses in the
latter case. One might further investigate this by a series of
MRI-simulations with initial vertical field of the form 
$B_z = B_0 \sin(2\pi x/L_x) + B_1$, where $|B_1| \ll |B_0|$, i.e., with a
controlled "spurious" field. Although from the physical point of view an exactly
vanishing vertical flux is rather a question of academic nature, the issue of
enhanced turbulent transport for net vertical field is still standing.

In our implementation we apply additional boundary conditions to the electric
field fluxes ${\bf G}$ to assure conservation of mean magnetic fields.  Due to
the different staggering of magnetic field components, which are face-centered,
the method has to be applied for the fluxes in all three directions. The
velocity offset $w$ enters the fluxes via the electromotive force:

\begin{equation}
{\bf E}(x,y,z) \mapsto {\bf E}(x\pm L_x, y\mp w t, z)\,
\pm w\;\hat{{\bf y}}\times {\bf B} \label{eq:map_E}
\end{equation}

For the $x$($z$)-direction only the $x$($z$)-component of the magnetic field
is needed to evaluate the additional second term. As the staggering of the
normal field component coincides with that of the fluxes no reconstruction is
needed in this case. Employing $B^E_{x\vert i,j,k} = B^W_{x\vert i+1,j,k}$
(respectively $B^T_{z\vert i,j,k} = B^B_{z\vert i,j,k+1}$) the characteristic
velocities cancel out and we yield

\begin{eqnarray}
{\bf G}^x_{i\phlf,j,k} &=& \widehat{\bf G}^x_{\itl\phlf,\jtl,k} \;\mp\; 
    w\;\hat{{\bf y}} B_{x\vert\itl\phlf,\jtl,k} \\
{\bf G}^z_{i,j,k\phlf} &=& \widehat{\bf G}^z_{\itl,\jtl,k\phlf} \;\mp\; 
    w\;\hat{{\bf y}} B_{z\vert\itl,\jtl,k\phlf}
\end{eqnarray}

for the numerical electric fluxes in $x$/$z$-direction. For the remaining
direction matters are a bit more complicated and we need the full
reconstruction to be consistent with the underlying numerical scheme:
\begin{equation}
{\bf G}^y_{i,j\phlf,k} = \widehat{\bf G}^y_{\itl,\jtl\phlf,k} \pm w\;\frac{
            b^{+}(B_x \hat{\bf x} + B_z \hat{\bf z})^N_{\itl,\jtl,k}
           -b^{-}(B_x \hat{\bf x} + B_z \hat{\bf z})^S_{\itl,\jtl,k}
            }{b^{+}-b^{-}}
\end{equation}

To conclude the discussion of the boundary conditions we want to remark that
the reconstruction of the magnetic field at the sheared interfaces preserves
the $\nabla\!\cdot\! {\overline{\bf B}}=0$ constraint to roundoff error (see
section~\ref{sec:adv_test}). The actual code
is publicly available on the internet at
\verb%www.aip.de/~gressel/index.php?id=code% and can be embedded into the
original NIRVANA package.

The current implementation supports distributed memory parallelism in the
form of blockwise domain decomposition along the z-coordinate. In this case all
the information needed to reconstruct the boundary values is available
locally, i.e., without MPI communications. Additional block distribution along
the x- and y-coordinates would in principle improve the surface to volume ratio
but would also add communication overhead. This is discussed in detail in
sections 4.3 and 4.4 of \cite{CK01}. Their Figure 5 shows that block
distribution along the x-coordinate is inefficient and performance gains for
additional distribution along the y-direction are moderate. They argue that
the poor performance for x-distribution might be a cache issue with the Fortran
column major ordering, which leaves this result marginally meaningful for our
implementation. In the case of vertically elongated boxes, like they are common
for simulations in galactic environments, distribution along z even outperforms
y- and y-z-distributions.

\subsection{Source terms}\label{sec:source} 

The second substantial ingredient of the shearingbox formalism are the source
terms in the so called Hill system. This approximation is based on the local
expansion of the equations of motion resulting in a tidal force $2q \Omega^2
x$ where $q=\d \ln \Omega/\d \ln R$ represents the (angular) shear parameter
for a differential rotation of the form $\Omega(R) \propto R^{-q}$. Together
with the Coriolis force the source terms for the momentum- and energy equation
are:

\begin{eqnarray}\label{eq:source_m}
S({\bf m}) & = & -2\varrho\Omega \hat{\bf z} \times {\bf v}
+2\varrho\, q\Omega^2x\hat{\bf x} = 
-2\varrho\Omega \hat{\bf z} \times \left( {\bf v} + q\Omega x \hat{\bf y}
\right) \label{eq:S_m}\\\label{eq:source_e}
S(e) & = & +2\varrho\Omega^2qx\hat{\bf x} \cdot {\bf v}.
\end{eqnarray}

Equation~(\ref{eq:S_m}) shows that the momentum source terms can be combined
and act as an effective\footnote{In contrast to the description in the
  Lagrangian frame of reference the angular velocity $\Omega=\Omega_0$ is not
  a function of radius here.} Coriolis force on the perturbed velocity $\delta
v_y=v_y+q\Omega x$. This would in principle allow for an exact Coriolis-update
in the form of an analytic rotation. Numerical tests, however, show that such
an update is not suitable for the multi-stage Runge-Kutta integration scheme.
Because operator-splitting is not favorable (for reasons described in
section~\ref{sec:epi} below) we decide to implement the source terms unsplit,
i.e., as explicit forces within the Runge-Kutta time integration scheme (see
eq. (\ref{eq:rk3})).

Gardiner \& Stone \cite{GS05} stress the importance of conserving the energy
contained in the \emph{epicyclic mode}. This ideally conserved quantity can be
derived from the energy budget in the limit of inviscid flow. The general
expression reads:

\begin{equation}
E_{\mathrm epi} = \half \varrho \left\langle u_R \right\rangle^2
+ \frac{(2\Omega)^2}{\kappa^2} \;
\half \varrho \left\langle u_\phi\right\rangle^2
= \half \varrho \left( \left\langle u_R \right\rangle^2 
+ \frac{2}{2-q} \; \left\langle u_{\phi}\right\rangle^2 \right),
\end{equation}

with $\kappa$ the epicyclic frequency. This formally looks like a kinetic
energy but also includes the potential energy with respect to the epicyclic
displacement. We have found that it is important to implement the source terms
in an unsplit fashion to avoid oscillations in this energy that would
otherwise arise from systematic splitting errors. This is discussed in more
detail in section~\ref{sec:epi} below.

\section{Test cases}\label{sec:tests}

We have validated our implementation with various simple advection tests, i.e.,
with source terms switched off, and have checked the mentioned conservation
properties. 

\subsection{Advection tests}\label{sec:adv_test}

\begin{table}\begin{center}
\begin{tabular}{|l|c|c|c|c|c|c|} \hline
resolution & & $24^2$ & $32^2$ & $48^2$ & $64^2$ & $96^2$ \\ \hline\hline
mass & a & 1.3371e-07 & 1.2407e-07 & 6.0259e-08 & 1.2674e-08 & 8.5488e-09 \\
\hline
     & b & 4.8018e-16 & 1.1181e-15 & 8.0027e-16 & 1.4405e-15 & 1.6005e-15 \\
\hline
\end{tabular}
\caption{Convergence study of the relative error in total mass: (a) without
special treatment, (b) with the method described above.}\label{tab:mass}
\end{center}\end{table}

As an example Table~\ref{tab:mass} shows the maximum relative error
in the total mass contained in our simulation box. Without special treatment of
the fluxes we find an error that decreases with increasing resolution, i.e., has
the form of a truncation error. With our modified treatment the total mass is
conserved to roundoff error. Albeit not explicitely shown in
Table~\ref{tab:mass}, this is also true for the momentum and total energy.

As discussed in section~\ref{sec:cons_magn}, for initially zero mean radial
field all components of the mean magnetic field are ideally conserved.
In Table~\ref{tab:magnetic} we show the components of the mean magnetic field,
normalized to corresponding rms values. One can see that without proper
mapping of the field fluxes the analytic constraint is only fullfilled to
roundoff error for $B_x$. With the method described above one can also conserve
the $B_z$ component to machine accuracy, while the error in the $B_y$
component can be reduced by about an order of magnitude. As can also be seen
from Table~\ref{tab:magnetic}, the maximum error in the solenoidal constraint
is of the order $10^{-13}$ while the average error is as low as
$5\times10^{-15}$. For the state of fully developed turbulence in the
MRI-simulations (see section~\ref{sec:mri}) the max./avg. values rise to
$10^{-10}$ and $10^{-13}$, respectively.

\begin{table}\begin{center}
\begin{tabular}{|l|c|c|c|c|c|} \hline
resolution & & $24^3$ & $32^3$ & $48^3$ & $64^3$ \\ \hline\hline
$\left \langle B_x \right \rangle/B_x^\mathrm{rms}$ 
 & a & 9.800e-16 & 1.062e-15 & 1.426e-15 & 5.285e-15 \\ \hline
 & b & 1.292e-15 & 1.512e-15 & 2.086e-15 & 5.658e-15 \\ \hline
$\left \langle B_y \right \rangle/B_y^\mathrm{rms}$ 
 & a & 1.043e-05 & 3.085e-06 & 4.021e-06 & 2.600e-06 \\ \hline
 & b & 2.454e-06 & 1.327e-06 & 4.000e-07 & 1.008e-07 \\ \hline
$\left \langle B_z \right \rangle/B_z^\mathrm{rms}$ 
 & a & 1.314e-05 & 1.602e-06 & 8.545e-06 & 7.646e-06 \\ \hline
 & b & 4.698e-17 & 4.574e-17 & 4.680e-17 & 7.304e-17 \\ \hline
max. $\nabla\!\cdot\!{\bf B}/|{\bf B}|$
 & a & 3.854e-13 & 3.448e-13 & 7.446e-13 & 5.671e-13 \\ \hline
 & b & 5.721e-13 & 1.879e-13 & 1.016e-12 & 1.177e-12 \\ \hline
avg. $\nabla\!\cdot\!{\bf B}/|{\bf B}|$
 & a & 5.493e-15 & 5.469e-15 & 4.575e-15 & 3.964e-15 \\ \hline
 & b & 5.157e-15 & 5.328e-15 & 4.658e-15 & 3.990e-15 \\ \hline

\end{tabular}
\caption{Convergence study of the relative error in the mean magnetic field
components (normalized to rms-values) and solenoidal constraint: (a) without
flux correction, (b) with
mapped fluxes.}\label{tab:magnetic}
\end{center}\end{table} 

In the following paragraph we analyze the epicyclic mode. As a real-life test
case we have also performed simple MRI simulations with initially vertical
magnetic field of zero net-flux.

\subsection{Conservation of epicyclic energy\label{sec:epi}}

By performing two-dimensional shearingsheet simulations the error in the
epicyclic energy is found to be independent of the mode amplitude. For constant
excitation amplitude the error is growing linearly in time. From the last row of
Table~\ref{tab:epi} one can see how the relative error per orbit decreases with
resolution, reflecting the third-order convergence of the underlying time
integration scheme. This is because due to the CFL stability condition, the
timestep is linearly proportional to the spatial resolution.

As already stated, our implementation of the source terms is integral part of
the third-order integration scheme. For comparison we have also tested
two conventional (operator split) methods for those terms: the first method
(A) is very similar to the one found in the ZEUS-code and directly (i.e.
forward Euler) integrates the Coriolis forces. The second method (B) treats
the Coriolis term analytically in form of a rotation of the momentum vector.
By expanding the trigonometric functions one can show that method A gives a
first order approximation to method B.

\begin{figure}
  \centering\includegraphics[width=8.5cm]{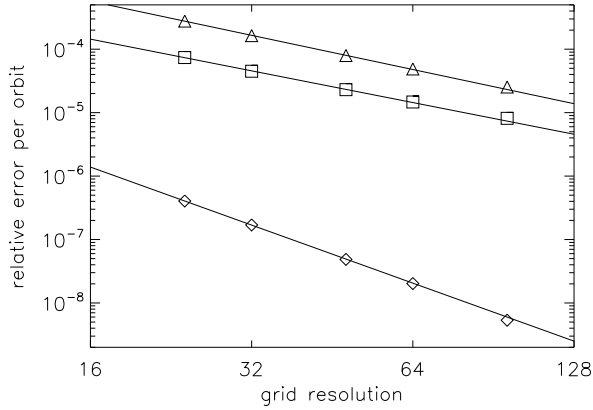}

\caption{Convergence of relative errors in epicyclic mode energy for method A
(triangles), method B (squares), and the unsplit method (diamonds). Lines
show least square fits with (logarithmic) slopes -1.79, -1.66, and
-3.04.}\label{fig:epi}
\end{figure}

Both methods (in contrast to the unsplit one) lead to oscillations, with
frequency $2\Omega$, in the epicyclic energy. We fit the resulting curves with
a function $f(t)=a t + b \sin(2 \Omega t)$. Table~\ref{tab:epi} shows the
oscillation amplitudes and linear growth-rates in the relative error as a
function of resolution. For reference we also include the fits for the
unsplit scheme which show negligible oscillations. Figure~\ref{fig:epi}
compares the rates of convergence for the described methods, clearly favoring
the unsplit approach.

\begin{table}\begin{center}
\begin{tabular}{|l|l|c|c|c|c|c|} \hline
resolution & $\,$ & $24^2$ & $32^2$ & $48^2$ & $64^2$ & $96^2$ \\ \hline\hline
method A   & a & 2.776e-04 & 1.637e-04 & 7.943e-05 & 4.877e-05 & 2.535e-05 \\
\hline
           & b & 0.032180  & 0.023989  & 0.015896  & 0.011886  & 0.007900  \\
\hline
method B   & a & 7.387e-05 & 4.497e-05 & 2.301e-05 & 1.473e-05 & 8.135e-06 \\
\hline
           & b & 0.013782  & 0.010275  & 0.006810  & 0.005093  & 0.003385  \\
\hline
unsplit    & a & 4.045e-07 & 1.695e-07 & 4.862e-08 & 2.023e-08 & 5.345e-09 \\
\hline
           & b & 1.285e-09 & 2.468e-10 & 8.284e-09 & 5.652e-09 & 4.505e-09 \\
\hline
           & c & 4.038e-07 & 1.695e-07 & 4.874e-08 & 1.747e-08 & 2.587e-09 \\
\hline
\end{tabular} 

\caption{Error growth-rates (a) and oscillation amplitudes
(b) of the operator-split methods A (direct integration of Coriolis-forces) and
B (exact rotation of momentum vectors). For the unsplit method we also show a
simple linear fit (c). }\label{tab:epi}
\end{center}\end{table}

\subsection{MRI with vertical field of zero net-flux}\label{sec:mri}

The most prominent application for the shearingbox model, of course, is the
magneto-rotational instability (MRI). We chose model parameters according to
previous simulations \cite{ZR00} with the old (non-conservative) NIRVANA-code,
that is very similar to the widely used ZEUS-code. For simplicity we neglect
stratification in this paper. A more sophisticated model, including
stratification and radiative cooling has been implemented and will be employed
in future work to explore MRI under conditions suitable to the interstellar
medium.

\begin{figure}
  \centering\includegraphics[width=12.5cm]{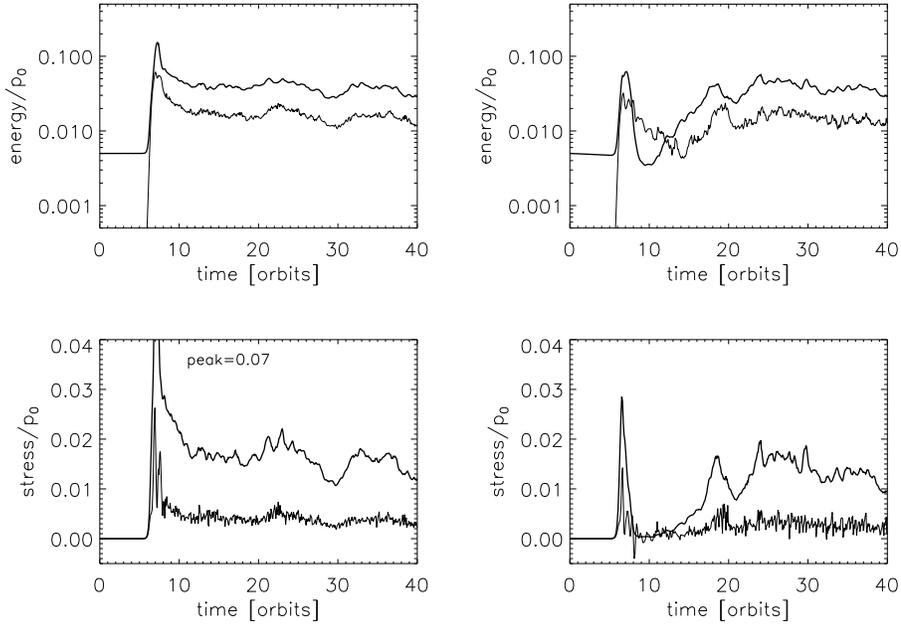}

\caption{Time evolution of volume averaged kinetic (thin lines) and magnetic
  (thick lines) energy density (top) and $R\phi$-components of Reynolds- and
  Maxwell-stresses (bottom) for the non-conservative scheme (left) and the
  conservative scheme (right).}\label{fig:MRI}
\end{figure} 

For now we use non-dimensional quantities, i.e., density is set to unity while
the pressure is $p=0.5 \times 10^{-6}$, such that the sound speed over the box
dimension matches the angular velocity $\Omega = 10^{-3}$. The initial
magnetic field is purely vertical and varies as ${\bf B}=B_0\sin(\pi
x/L_x)\,\hat{\bf z}$, resulting in a vanishing net vertical flux. The field
amplitude $B_0=1.121 \times 10^{-7}$ corresponds to a plasma parameter of
$\beta=100$ at the peaks of the sine-profile.

\begin{figure}
  \centering\includegraphics[bb=40 0 480 160, width=12.5cm]{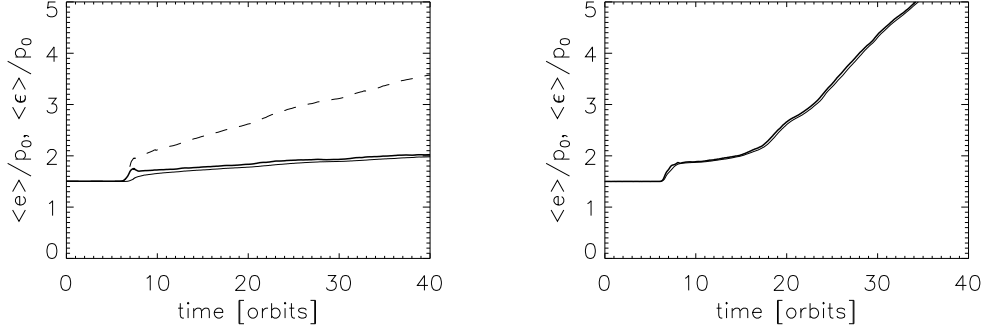}

\caption{Time evolution of volume averaged total (thick lines) and
  thermal (thin lines) energy for the non-conservative scheme (left) and the
  conservative scheme (right). For the former we also plot the work done by
  the boundary conditions (dashed line). All quantities are normalized by the
  initial gas pressure $p_0$.}\label{fig:MRI-heating}
\end{figure}

We apply a box geometry of $[-0.5,0.5]\times[0,4] \times[-2,2]$ with a
resolution of 64x128x128 grid cells and perform computations on the new and
the old code version to compare the conservative vs. the non-conservative
scheme.  The results are directly compared in Figure~\ref{fig:MRI}. The
saturated stresses and energies are of comparable magnitude although the
initial growth of the instability and the breakdown of the channel-solution
into chaotic turbulence differs quite a bit. The breakdown of the linear
solution is related to parasitic instabilities that seem to be resolved
differently in the two codes. The weaker initial peak for the conservative
scheme is consistent with the results in \cite{GS05}. Saturation amplitudes
are compatible with earlier results \cite{HGB95,ZR00,GS05}.

Although the codes compare quite well with respect to the kinematic
quantities (see Figure~\ref{fig:MRI}), there is a major difference concerning
the thermalization of the extracted turbulent energy. As can be seen in the
right panel of Figure~\ref{fig:MRI-heating} the conservative code efficiently
transfers the released energy into thermal energy, which leads to a rapid
heating of the gas.  The non-conservative code shows much less heating. Compared
to the work exerted by the boundary-stresses\footnote{This quantity is obtained
by accumulating the rhs of Eq.~(\ref{eq:ene_cons}).}, that is shown as a
dashed line (on top of the total energy) in the left panel of
Figure~\ref{fig:MRI-heating}, there is a considerable amount of energy lost.

\section{Conclusions}

This paper described the implementation of a shearingbox environment for the
flux-conservative/constraint-transport MHD code NIRVANA. To our knowledge
this is the first time such an implementation has been developed for a
central-type numerical scheme. We showed that shift-periodic boundary conditions
accounting for a prescribed linear shear flow across the computational domain
can be handled within a conservative framework.  This comes about by a proper
modification of the hydrodynamical numerical fluxes with the mappings of
$y$-momentum flux and total energy flux solely expressed in terms of known
mapped fluxes and the velocity offset of the background shear flow. Such
mappings are, in principle, independent of the underlying numerical scheme,
provided it is given in flux-conservation form.

In contrast, our proposed mappings of the (face-centered) electric field
fluxes contain explicit information on the magnetic field components at the
boundary as well as on the characteristic speeds used by the numerical scheme.
Nevertheless, we demonstrated by numerical experiment that such a mapping is
divergence-free, exactely conserves the $x$- and $z$-component
of mean magnetic field and the corresponding $y$-component up to
truncation error but, obviously, is scheme-dependent. An alternative method
would be to directly map the edge-centered electric field fluxes, as those are
the quantities which enter the constraint-transport equations. However, initial
investigations of that approach have not been proven successful to date.

\begin{ack}
The MRI-computations were performed on the Sanssouci PC-cluster of the
Astrophysical Institute Potsdam. This work was supported by Deutsche
For\-schungs\-ge\-mein\-schaft (DFG).
\end{ack}

% \appendix

% \section{}
% \label{}

\end{document}